\documentclass[aps,twocolumn,pre,showkeys]{revtex4}

\usepackage{graphpap}
\usepackage[dvips]{graphicx}
\usepackage[dvips]{graphics}
\usepackage{color}
\usepackage{hyperref}
\usepackage{amsmath}

\graphicspath{{./figures/}}

\usepackage{nomencl}
\makenomenclature

\begin{document}

\title{Coupling graphene mechanical resonators to superconducting microwave cavities} 

\author{P. Weber\footnotemark[1], J. G\"uttinger\footnotemark[1]\footnotemark[2], I. Tsioutsios, D. E. Chang, A. Bachtold}

\affiliation{ICFO-Institut de Ciencies Fotoniques,Mediterranean Technology Park, 08860 Castelldefels (Barcelona), Spain}


\begin{abstract}
Graphene is an attractive material for nanomechanical devices 
because it allows for exceptional properties, such as 
 high frequencies and quality factors, and low mass. 
An outstanding challenge, however, has been to obtain large coupling between the motion and external systems for efficient readout and manipulation. Here, we report on a novel approach, in which we capacitively couple a high-Q graphene mechanical resonator ($Q \sim 10^5$) to a superconducting microwave cavity. The initial devices exhibit a large single-photon coupling of $\sim 10$~Hz. Remarkably, we can electrostatically change the graphene equilibrium position and thereby tune the single photon coupling, the mechanical resonance frequency and the sign and magnitude of the observed Duffing nonlinearity. The strong tunability opens up new possibilities, such as the tuning of the optomechanical coupling strength on a time scale faster than the inverse of the cavity linewidth. With realistic improvements, 
it should be possible to enter the regime of quantum optomechanics. 
\end{abstract}


\maketitle

\footnotetext[1]{
These authors contributed equally to this work.
}
\footnotetext[2]{Corresponding author. E-mail johannes.guettinger@icfo.es.}

Mechanical resonators based on individual nanotubes and graphene flakes have outstanding properties. Their masses are ultra-low, their quality factors can be remarkably high, the resonance frequencies are widely tunable, and their equilibrium positions can be varied by a large amount. As a result, the resonators can be used as sensors of mass~\cite{chiu2008, chaste2012} and force~\cite{bunch2007,moser2013,stapfner2013} with unprecedented sensitivities, and they can be employed as parametric amplifiers~\cite{eichler11para} and as tunable oscillators~\cite{ayari07,eichler11para,chen2013osc}. Thus far, all these scientific applications are accomplished in the classical regime.

Reaching the quantum regime with mechanical resonators has attracted considerable interest~\cite{poot2012,aspelmeyer2013}. Thus far, three groups have been successful in this quest by demonstrating that the number of vibrational quanta can be lowered below one~\cite{oconn2010,teufel2011b,chan2011}. These three groups were using different resonators, namely, a piezoelectric resonator, a superconducting resonator, and an opto-mechanical crystal. 
There is now an intense effort from the community to develop new types of opto-mechanical and electro-mechanical devices, the goal being to explore new scientific and technological applications when these devices will enter the quantum regime.
This includes levitating particles~\cite{chang10,kiesel2013,gieseler2013}, optically trapped cantilevers~\cite{norte2012}, and heavy pillars~\cite{kuhn2011} to test the foundations of quantum mechanics; metal coated silicon nitride membranes to coherently convert radio-frequency photons to visible photons~\cite{bagci2013,andrews2013}; microdisks and nanopillars to boost the single-photon coupling and to enter the ultra strong coupling regime~\cite{ding2010,yeo2013}. In this context, the unique properties of nanotube and graphene resonators are very interesting.     

Although nanotubes and graphene have exceptional properties, an outstanding challenge in approaching the quantum regime has been the development of efficient coupling to external elements, which would enable motional readout and manipulation. For example, while graphene has been coupled to an optical cavity~\cite{barton12}, the 2.3\% optical absorption of graphene makes it extremely challenging to reach the quantum regime, due to heating of the graphene and quenching of the optical cavity finesse. Here, we employ a different strategy, which is to couple the mechanical resonator capacitively to a superconducting cavity~\cite{regal08,hertzberg09,roch10,teufel2011b,massel2011,zhou2013}. This is a promising approach with graphene resonators, because the two-dimensional shape of graphene is ideal for large capacitive coupling. 

In this work we report on the integration of a 
circular graphene resonator 
with a superconducting microwave cavity. 
We use a transfer technique to precisely 
position a high-quality exfoliated graphene flake with respect to a predefined superconducting cavity. 
We develop a reliable method to reduce the separation between the graphene
membrane and the cavity
by tightly clamping the graphene sheet in between a support electrode and a cross-linked Polymethyl methacrylate (PMMA) structure.
We show that this technique allows us to 
improve the mechanical stability and to achieve high mechanical quality factors. 
By pumping the cavity on a motional sideband, we are able to sensitively readout the graphene motion. 
Importantly, by applying a constant voltage $V_\mathrm{g}^\mathrm{DC}$ to the graphene, 
the properties of the optomechanical device can be dramatically tuned. 
Namely, large static forces can be produced, allowing to tune the steady-state displacement, the mechanical resonance frequency, the optomechanical coupling, and the mechanical nonlinearities. Such a tunability cannot be achieved in other opto-mechanical systems.

\begin{figure*}[t]
\includegraphics[width=\textwidth]{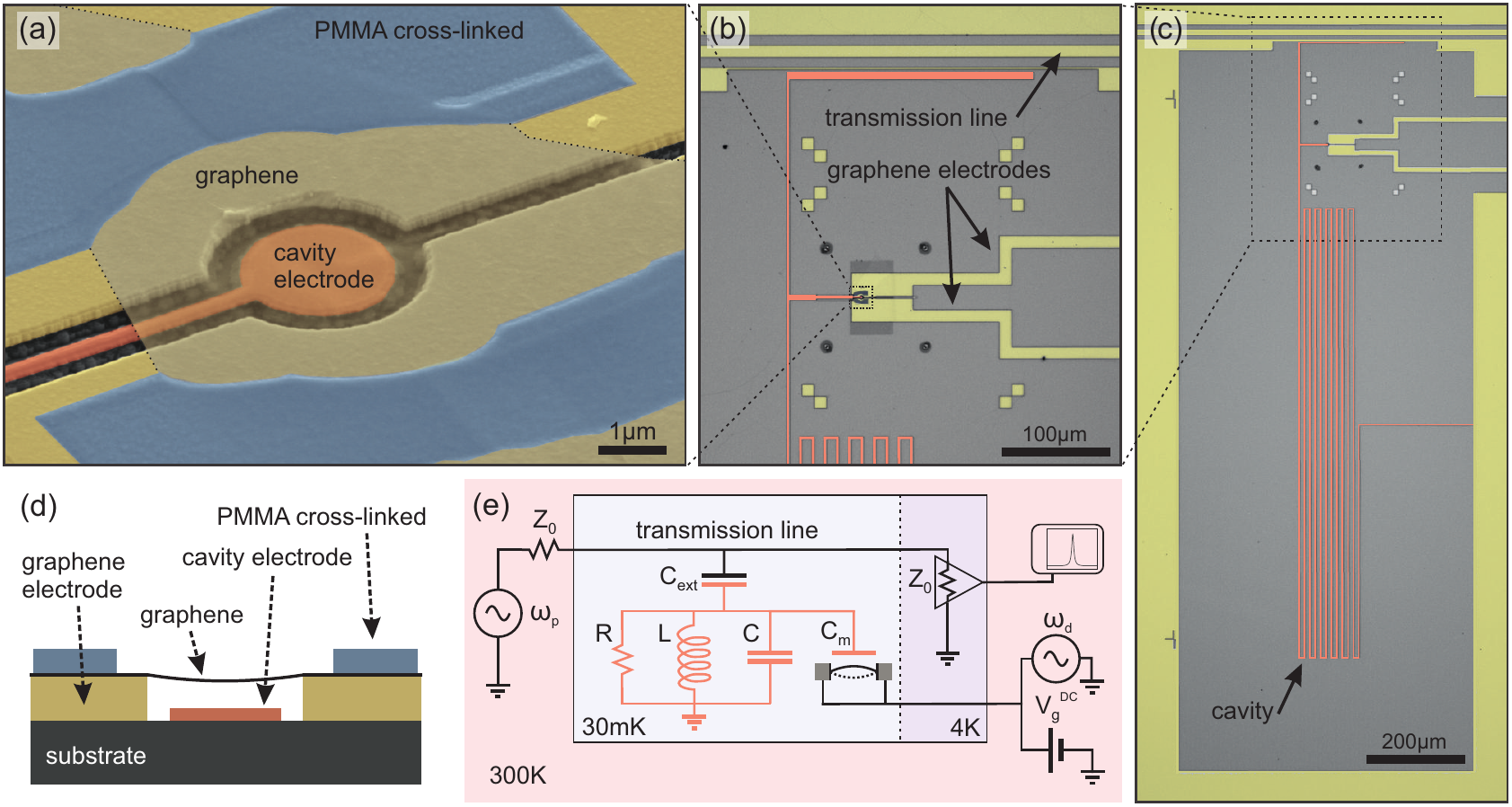}
\caption{(a) False color SEM image of a circular graphene resonator capacitively coupled to a cavity electrode. The graphene sheet is clamped in between cross-linked PMMA and graphene support electrodes. (b,c) Optical microscope images of the superconducting cavity, two electrodes contacting the graphene flake, and a capacitively coupled transmission line. (d) Schematic cross-section of the mechanical resonator and the cavity counter electrode.
(e) Schematic of the measurement circuit. A coherent pump field at $\omega_\mathrm{p}$ is applied to the transmission line. The graphene mechanical resonator is driven by a field at $\omega_\mathrm{d}$ and a constant voltage $V_\mathrm{g}^\mathrm{DC}$. The microwave signal from the cavity is amplified at 4~K with a HEMT amplifier and recorded at room temperature with a spectrum analyzer. The impedance $Z_0$ is $50~\Omega$.
}
\label{f:figure 1}
\end{figure*} 

Our device (see Fig.~1a-d) consists of a superconducting microwave cavity, modeled as a $LC$-circuit with angular frequency $\omega_\mathrm{c}=1/\sqrt{LC_\mathrm{tot}}\approx 6.7~$GHz,  capacitance $C_\mathrm{tot} \approx 90~$fF, inductance $L \approx 6.3~$nH, and characteristic impedance $Z_\mathrm{c} = \sqrt{L/C_\mathrm{tot}} \approx 260~\Omega$. The total capacitance $C_\mathrm{tot}=C+C_\mathrm{ext}+C_\mathrm{m}(z)$ effectively consists of a cavity capacitance $C\approx 85$~fF, a contribution $C_\mathrm{ext}\approx 5$~fF from the external feedline, and importantly, a contribution $C_\mathrm{m}(z) \approx 0.3-0.4$~fF that depends on the graphene position $z$, which arises from the graphene acting as a moving capacitor plate. 
A small displacement $z$ therefore produces a shift in $\omega_\mathrm{c}$ quantified by the optomechanical coupling $G_0 = \frac{\partial \omega_\mathrm{c}}{\partial z}$.
As a result, the interaction between the mechanical resonator and the superconducting cavity can be described by the Hamiltonian $H_\mathrm{int} = \hbar G_0 n_\mathrm{p} z$~\cite{teufel2011b} with $n_\mathrm{p}$ the number of pump photons in the cavity. 
The characteristic coupling at the level of the zero-point motion $z_\mathrm{zp} = \sqrt{\hbar/2m_\mathrm{eff}\omega_m}$
is given by the so-called single-photon coupling $g_0 = G_0 z_\mathrm{zp}$,
with $m_\mathrm{eff}$ the effective mass and $\omega_\mathrm{m}/2\pi$ the resonance frequency of the 
mechanical mode of interest. Central to this work is \textit{(i)} that the low mass of graphene boosts  $z_\mathrm{zp}$ and thus $g_0$, and \textit{(ii)} that $C_\mathrm{m}$ and $g_0$ can be tuned electrostatically with $V_\mathrm{g}^\mathrm{DC}$.

We start with engineering considerations in order to maximize the coupling  $g_0$. 
When describing $C_\mathrm{m}$ by a plate capacitor and noting that $C \gg C_\mathrm{ext} \gg C_\mathrm{m}(z)$ in our device, we have $g_0 \approx  \frac{\omega_\mathrm{c}}{2 C}\frac{\partial C_\mathrm{m}(z) }{\partial z} z_\mathrm{zp} \propto \sqrt{\frac{A}{\omega_\mathrm{m}}}\frac{\omega_\mathrm{c}}{C d^2}$ using $\frac{\partial C_\mathrm{m}(z) }{\partial z} \propto A/d^2$ and $z_\mathrm{zp} \propto 1/\sqrt{A \omega_\mathrm{m}}$. Here $A$ is the area of the suspended graphene region and $d$ is the separation between the graphene membrane 
and its cavity counter electrode. In order to optimize the coupling $g_0$, it is crucial to minimize both $C$ and $d$. 
To this end, we utilize a narrow cavity conductor structured in a meander to increase $L$, while minimizing the capacitance to the ground 
for a given $\omega_\mathrm{c}$. 
In order to be able to tune $d$ with $V_\mathrm{g}^\mathrm{DC}$, we use a cavity that is shorted to ground on one side, allowing for a well defined electrical DC potential. The fundamental mode of the cavity is a quarter wavelength standing wave, with a voltage node at the shorted end and the largest voltage oscillation amplitudes at the open end. 
The graphene membrane is coupled close to the open end of the cavity to harness the largest cavity fields
(see Fig.~1b,c)~\cite{day2003,regal08,teufel2009}. 
Using this geometry, we achieve a cavity capacitance of $C \approx 90$~fF.
This compares favorably with $C = 18$~fF-1~pF in previous studies~\cite{regal08,hertzberg09,roch10,teufel2011b,massel2011,zhou2013}. 
Note that the lowest values for $C$ have been achieved in closed-loop cavities, where the mechanical capacitance is incorporated between the two ends of a half-wavelength cavity~\cite{teufel2011b,massel2011}. In this geometry the two electrodes of the mechanical capacitance are shorted over the cavity, so that no static DC potential can be applied. 
Compared to the capacitance of a gated half-wavelength cavity~\cite{delbecq11,frey12,petersson2012}, the capacitance of a quarter wavelength cavity is lowered by a factor of two. 

In order to detect the vibrations of the graphene resonator, we couple the open-end of the superconducting cavity to a microwave transmission line through the capacitance $C_\mathrm{ext}$.  
The transmission line is used to pump the superconducting cavity at frequency $\omega_\mathrm{p}/2\pi$ with input power $P_\mathrm{p,in}$. The transmission line is also employed to measure the output power $P_\mathrm{out}$ of the cavity at frequency $\omega_\mathrm{c}/2\pi$. $P_\mathrm{out}$ is amplified at 4~K by a high-electron-mobility transistor (HEMT) with a noise temperature of about $2$~K and measured in a spectrum analyzer (see schematic in Fig.~1e and SI).

We use a graphene resonator with a circular shape. This geometry improves the attachment of the graphene sheet to its support when compared to the doubly-clamped resonator geometry. As further discussed below, a strong attachment of the graphene to its support is crucial to be able to lower $d$.
Another advantage of circular graphene resonators over doubly-clamped resonators is that the quality factor tends to be larger~\cite{barton11}.
In addition, the mechanical eigenmodes of circular resonators are well defined~\cite{barton11,eriksson2013}. In particular, it avoids the formation of modes localized at the edges, which were observed in doubly-clamped resonators~\cite{garcia08}. 

\begin{figure}[hbt]
\centering
\includegraphics[width=\linewidth]{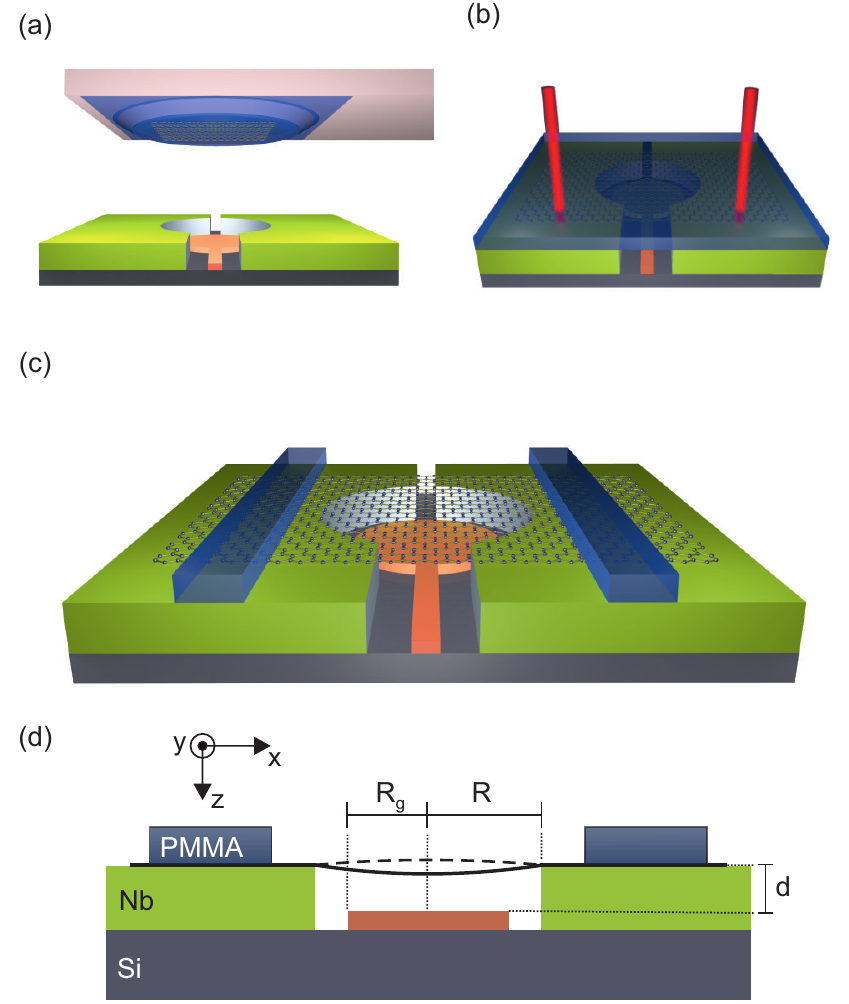}
\caption[Figure 2] {Fabrication process for PMMA-clamped graphene mechanical resonator. (a) Transfer of graphene with PMMA (blue) onto predefined structure (yellow/green, gray). (b) Cross-linking part of the transferred PMMA by electron-beam overexposure (red). (c) Schematic of the final device. (d) Cross-section of the device.}
\label{f:figure2}
\end{figure}

To fabricate the devices, we start by carving out the superconducting cavity structure from a 200~nm thick sputtered Niobium (Nb) film by ion milling and reactive ion etching (see SI). 
We employ a PMMA supported transfer technique pioneered at Columbia~\cite{dean10} to position graphene flakes on the superconducting cavity structure. For this, we exfoliate graphene sheets from large graphite crystals onto a silicon (Si) chip covered by a polymer film consisting of 100~nm polyvinyl alcohol (PVA) and 200~nm PMMA 495K. The thickness of the PVA/PMMA film is optimized to give the largest optical contrast of graphene flakes in an optical microscope. In particular, it allows to calibrate the number of layers of the graphene flake~\cite{blake07,wang07}. The solvability of PVA in water is used to separate the Si chip from the PMMA with the graphene. Using a brass slide with a volcano-shaped hole, the membrane is fished from the water and dried on a hotplate. When drying, the PMMA membrane gets uniformly stretched across the volcano hole. 
By mounting the slide upside down into a micromanipulator, the graphene sheet can be aligned and transferred onto the pre-patterned superconducting cavity structure, as illustrated in Fig.~2a.
To improve the attachment of the graphene flake to its support, it was shown that it is important to clamp the graphene membrane on the two sides of its surface~\cite{bao12}.
For this, we crosslink part of the transferred PMMA with a $10,000~\mu$C/cm$^2$ electron beam dose (Fig.~2b). The unexposed PMMA is removed in $80^\circ$C hot N-Methyl-2-pyrrolidone (NMP), followed by critical point drying of the device. 
As a result, the graphene is firmly sandwiched between the support electrode and the crosslinked PMMA (Fig.~2c,d).
Using this technique the graphene sheet is less likely to collapse against its counter electrode. This allows us to increase the success yield of the device fabrication. We have successfully lowered the separation to $d = 85$~nm for a $3.5~\mu$m diameter graphene resonator, which is equal to the best diameter-separation ratio of $2R/d = 40$ reported so far for graphene resonators~\cite{lee13}. In addition, the strong attachment between the graphene and its support allows us to electrostatically tune the equilibrium position by a large amount (see below).  

In this letter we present results measured at $30~$mK for two different graphene devices, hereafter called devices $A$ and $B$. Device $A$ is a three layer graphene resonator with radius $R = 1.75~\mu$m and with $d=95~$nm. The number of layers is determined from optical contrast measurements~\cite{blake07,wang07}. The radius of the counter electrode is $R_\mathrm{g} = 1.1~\mu$m (see Fig.~2d). Device $B$ is a four layer graphene resonator with the same membrane radius, $d=135~$nm and $R_\mathrm{g} = 1.25~\mu$m.

The principle of mechanical vibration readout is analogous to Stokes and anti-Stokes Raman scattering. By pumping the cavity at $\omega_\mathrm{p}$, sidebands in energy are created at $\omega_\mathrm{p} \pm \omega_\mathrm{m}$ due to the coupling of the photons with the mechanical motion. 
If the pump is detuned such that the upper sideband frequency is matched with the cavity resonance frequency $\omega_\mathrm{c} = \omega_\mathrm{p} + \omega_\mathrm{m}$ (see Fig.~3a), the anti-Stokes scattering is resonantly enhanced. 
Then, the rate of the anti-Stokes scattering per phonon is given by $\Gamma_\mathrm{opt} \approx 4 n_\mathrm{p} g_0^2/\kappa$, 
with  $n_\mathrm{p} \propto P_\mathrm{p,in}(\omega_\mathrm{p})$ the number of photons in the cavity.
We drive the graphene resonator by applying a constant voltage $V_\mathrm{g}^\mathrm{DC}$ and an oscillating voltage with amplitude $V_\mathrm{g}^\mathrm{AC}$ at a frequency $\omega_\mathrm{d}/2\pi$ close to $\omega_\mathrm{m}/2\pi$ so that $\omega_\mathrm{d} = \omega_\mathrm{c}-\omega_\mathrm{p}$. As a result, the graphene resonator vibrates at $z(t) = \hat{z}\cos{(\omega_\mathrm{d}t+\phi)}$ with $\phi$ the phase difference between the displacement and the driving force. The output power at $\omega_\mathrm{c}$ is
\begin{equation}
P_\mathrm{out} =P_\mathrm{p,in} \frac{\kappa_{ext}^2}{\kappa^2+4(\omega_\mathrm{c}-\omega_\mathrm{p})^2} 4 \frac{g_0^2}{\kappa^2} \frac{\left\langle z(t)^2\right\rangle}{2z_\mathrm{zp}^2}.
\label{eq:Pout}
\end{equation}  
From a transmission measurement of the feedline we readily get the resonance frequency of the cavity $\omega_\mathrm{c} /2\pi = 6.73~$GHz and the total linewidth $\kappa/2\pi =  \kappa_\mathrm{ext}/2\pi + \kappa_\mathrm{int}/2\pi = 15.2~$MHz with $\kappa_\mathrm{ext}/2\pi = 2$~MHz the coupling rate of the superconducting cavity to the feedline and $\kappa_\mathrm{int}/2\pi = 13.2$~MHz the internal loss rate of the cavity.
A detailed analysis of the circuit, which includes a resistance to describe the losses in the graphene flake and the DC connections, shows that this additional resistance contributes roughly 20\% to $\kappa_\mathrm{int}$ (see SI). The high value of $\kappa_\mathrm{int}$  is attributed to the contamination and imperfections of the cavities. Indeed, we tested the cavity of devices $A$ and $B$ at $T = 4.2$~K before the transfer of the graphene flakes, and we observed larger $\kappa_\mathrm{int}$ than what we usually observe in devices processed in the same way. 

\begin{figure}[htb]
\centering
\includegraphics[width=\linewidth]{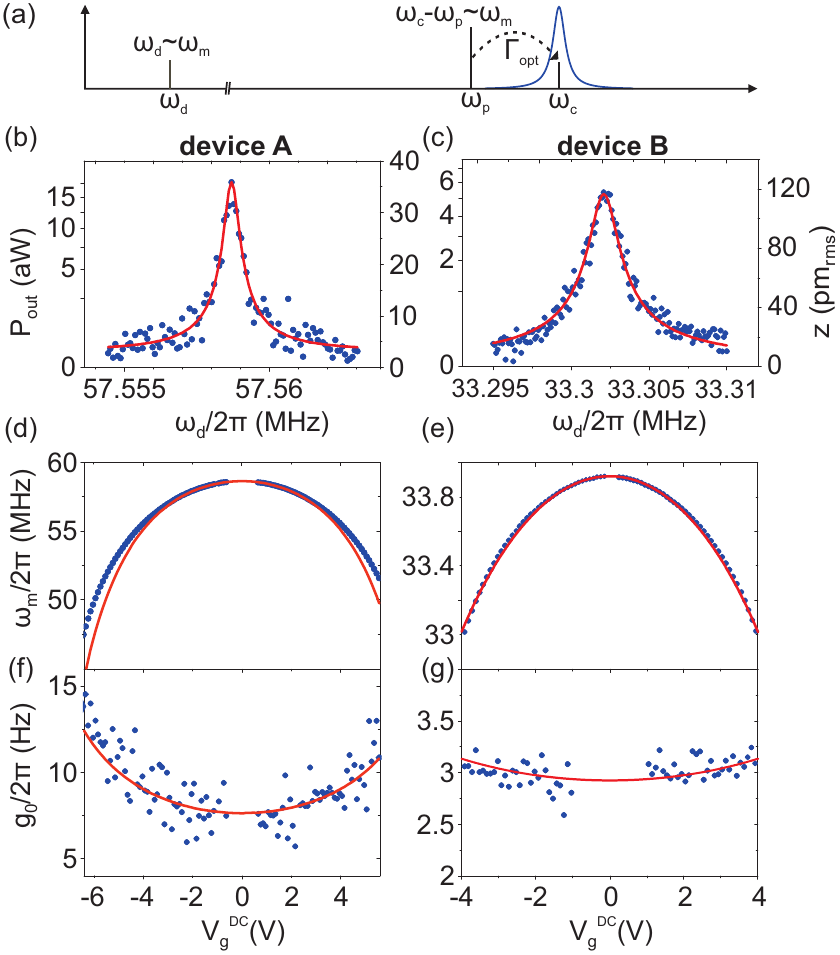}
\caption[Figure 3] {(a) Measurement scheme: If the pump frequency is detuned such that $\omega_\mathrm{p} = \omega_\mathrm{c} - \omega_\mathrm{m}$, anti-Stokes scattering with phonons at rate $\Gamma_\mathrm{opt}$ leads to a detectable photon population at $\omega_\mathrm{c}$. (b),(c) Sideband measurement of the mechanical motion for device $A$ with $V_\mathrm{g}^\mathrm{DC}=-2.894~$V and $V_\mathrm{g}^\mathrm{AC}=190~$nV, and for device $B$ with $V_\mathrm{g}^\mathrm{DC}=3.405~$V and $V_\mathrm{g}^\mathrm{AC}=4.3~\mu$V. Red lines are Lorentzian fits to the data which yield a mechanical quality factor of $Q_\mathrm{m} = 100,000$ in device $A$ and $Q_\mathrm{m} = 17,700$ in device $B$. The calculated motional rms amplitude $z$ is plotted on the right scale. (d),(e) Mechanical resonance frequency as a function of $V_\mathrm{g}^\mathrm{DC}$. We have compensated $V_\mathrm{g}^\mathrm{DC}$ by an offset of 0.434~V for device $A$ and 0.395~V for device $B$. 
In addition to capacitive softening, the static deflection $z_{\mathrm{s}}$ of the resonator towards the cavity counter electrode is considered in order to account for the measurement (red line).
(f),(g) Single-photon coupling rate $g_0 = G_0 z_\mathrm{zp}$. By including the static displacement $z_\mathrm{s}$ we are able to model the single-photon coupling as a function of $V_\mathrm{g}^\mathrm{DC}$ (red line). 
}
\label{f:figure 3}
\end{figure}
Figures~3b,c show the resonance of the driven vibrations for the fundamental modes of device $A$ and $B$. Modes at higher frequencies are observed as well, but they are hardly detectable. For device $A$ we extract the mechanical quality factor $Q_\mathrm{m} = \omega_\mathrm{m}/\gamma_\mathrm{m} \approx  100,000$ from the linewidth of the resonance $\gamma_\mathrm{m}/2\pi = 575~$Hz. This $Q_\mathrm{m}$ is comparable to the largest values reported thus far for graphene resonators~\cite{eichler11}, showing that our fabrication process does provide us with mechanical resonators of excellent quality. We used $n_\mathrm{p} = 8000$ photons for this measurement, so that $\Gamma_\mathrm{opt}/2\pi \approx 0.12~$Hz. With these parameters, the measurement imprecision, estimated to be $2.5~\mathrm{pm}/\sqrt{\mathrm{Hz}}$, is limited by the noise of the low-temperature HEMT amplifier. For comparison, the height of the resonance in the power spectral density of the thermal motion at 30~mK is $(7~\mathrm{fm})^2/$Hz (see SI). 
In device $B$ we measure a quality factor of $Q_\mathrm{m} =  17,700$.
We attribute this lower $Q_\mathrm{m}$ to the fact that the device was imaged in a scanning electron microscope (SEM) before the measurements, where the graphene surface got contaminated by amorphous carbon. This measurement was done with $n_\mathrm{p} = 4500$ photons, corresponding to $\Gamma_\mathrm{opt}/2\pi \approx 0.01~$Hz. If we further increase the pump power we observe a reduction of the quality factor. We attribute this reduction of $Q_\mathrm{m}$ to Joule heating in the graphene flake. A rough estimate of the heating can be made by measuring the quality factor as a function of the temperature of the cryostat. From this comparison, $n_\mathrm{p} = 10^6$ corresponds for instance to a temperature of about 200~mK (see SI).

The resonance frequency decreases upon increasing $|V_\mathrm{g}^\mathrm{DC}|$ (see Fig.~3d,e). This reduction of the resonance frequency has been observed previously in graphene resonators under tension~\cite{chen2009,eichler11,song2011}. 
This softening of the resonator is attributed to the change of the restoring potential of the resonator by the capacitive energy~\cite{koz06,chen2009,eichler11,song2011}. We model the mechanical resonator with a circular membrane under tension~\cite{landau1970} to quantify the observed dependence. When neglecting static deflection, the frequency dependence is given by
\begin{equation}
\omega_\mathrm{m}(V_\mathrm{g}^\mathrm{DC}) = \sqrt{\frac{4.92 Eh \epsilon}{m_\mathrm{eff}} - \frac{0.271}{m_\mathrm{eff}}\frac{\epsilon_0 \pi R_\mathrm{g}^2}{d^3} (V_\mathrm{g}^\mathrm{DC})^2},
\label{eq:omegam}
\end{equation}
with $\epsilon$ the strain in the graphene sheet, $E \approx 1$~TPa the Young's modulus of graphite, $h= n_\mathrm{g}\times 0.34~$nm the graphene thickness, $n_\mathrm{g}$ the number of graphene layers~\cite{bunch2007,chen2009} and $m_\mathrm{eff} = 0.27\pi R^2\rho_\mathrm{2D}$ the effective mass of the fundamental mode (see SI). The two dimensional mass density $\rho_\mathrm{2D} = \eta n_\mathrm{g} \rho_\mathrm{graphene}$ includes the graphene mass density $\rho_\mathrm{graphene} = 7.6\times10^{-19}~$kg$/\mu$m$^2$ and a correction factor $\eta \geq 1$ to account for contamination on the graphene surface. From a fit of Eq.~\eqref{eq:omegam} to the measurements around $V_\mathrm{g}^\mathrm{DC} = 0$ (in Fig.~3d,e), we extract $m_\mathrm{eff} = 13\cdot 10^{-18}~$kg and $\epsilon= 0.036\%$ for device $A$, and $m_\mathrm{eff} = 36\cdot 10^{-18}~$kg and $\epsilon= 0.024\%$ for device $B$. The obtained mass is $\eta = 2.2$ times larger than the total graphene mass for device $A$ and $\eta =4.5$ times larger for device $B$.
The larger $\eta$ for device $B$ might be attributed to the amorphous carbon deposited during SEM inspection.
The tension is intermediate compared to previous measurements, 
where $\epsilon$ ranges from $0.002\%$ to $1\%$~\cite{bunch2007,barton11,eichler11,chen13}. 
\begin{figure}[htb]
\centering
\includegraphics[width=\linewidth]{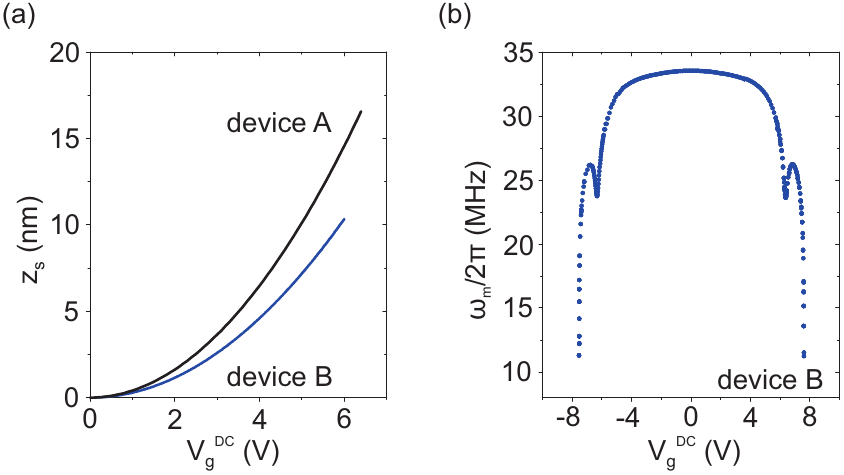}
\caption[Figure 4] {(a) Static displacement of the center of the membrane calculated from Eq.~\eqref{eq:staticdisp} with constants $c_{\mathrm{s}} = 0.405~$nm/V$^2$ for device $A$ and $c_{\mathrm{s}} = 0.287~$nm/V$^2$ for device $B$. (b) Mechanical resonance frequency as a function of $V_\mathrm{g}^\mathrm{DC}$ for device $B$.}
\label{f:figure 4}
\end{figure}

In order to account for the variation of $\omega_\mathrm{m}$ for large $V_\mathrm{g}^\mathrm{DC}$ in Fig.~3d,e, the static deflection of the graphene sheet towards the cavity counter electrode has to be considered.
The static displacement of the center of the membrane $z_\mathrm{s}$ is given by 
\begin{equation}
z_\mathrm{s} = \frac{\epsilon_0 R_\mathrm{g}^2}{8Eh\epsilon d^2}(V_\mathrm{g}^\mathrm{DC})^2 = c_\mathrm{s} (V_\mathrm{g}^\mathrm{DC})^2
\label{eq:staticdisp}
\end{equation}
for small displacement compared to $d$. Although the renormalization of the mechanical frequency due to static displacement cannot be solved exactly, as an approximation we can include $z_\mathrm{s}$ in Eq.~\eqref{eq:omegam} using $d = d_0- z_\mathrm{s}$, with $d_0$ the separation for $V_\mathrm{g}^\mathrm{DC} = 0$.
We get a good agreement for $\omega_\mathrm{m}(V_\mathrm{g}^\mathrm{DC})$ between the measurements and theory without any fitting parameter over the $V_\mathrm{g}^\mathrm{DC}$ range shown in Fig.~3d,e. The effect of $z_\mathrm{s}$ on the shift in $\omega_\mathrm{m}$ is 42\% at $V_\mathrm{g}^\mathrm{DC} = -6$~V for device $A$ and 10\% at $V_\mathrm{g}^\mathrm{DC} = 4$~V for device $B$. 
The expected variation of $z_\mathrm{s}$ is plotted in Fig.~4a.

The softening of the graphene resonator becomes enormous upon further increasing $V_\mathrm{g}^\mathrm{DC}$, with a reduction of $\omega_\mathrm{m}$ by a factor of three down to $\approx 10$~MHz as shown in Fig.~4b for device $B$. This reduction of $\omega_\mathrm{m}$ is large compared to that measured in previous works~\cite{koz06,chen2009,song2011}. 
Such a large reduction is expected when the capacitive force becomes comparable to the restoring force of the resonator. When the two forces are equal, $\omega_\mathrm{m}$ drops to zero and the resonator collapses against the counter electrode~\cite{sillanpae11}. Even though further work is needed to understand the quantitative dependence of $\omega_\mathrm{m}$ on $V_\mathrm{g}^\mathrm{DC}$, it reveals that the graphene resonators we fabricate can bend by a large amount without being ripped apart due to the large induced strain and without sliding with respect to the anchor electrodes.

The static displacement of the graphene sheet also changes the resonance frequency of the microwave cavity upon varying $V_\mathrm{g}^\mathrm{DC}$. As the graphene moves closer to the cavity counter electrode, the total capacitance of the cavity increases, so that the cavity frequency decreases. For $\Delta V_\mathrm{g}^\mathrm{DC}=6~$V the decrease is $\Delta \omega_\mathrm{c}/2\pi=2~$MHz in device $A$. The measured $\Delta \omega_\mathrm{c}$ agrees well with the shift expected from the static displacement (see SI).  

Our device layout allows us to get 
large couplings $g_0$ between the mechanical resonator and the superconducting cavity
(Fig.~3f,g). We extract $g_0$ from the measurements of the response of driven vibrations at $\omega_\mathrm{d} =  \omega_\mathrm{m}$ using Eq.~\eqref{eq:Pout} where $\left\langle z(t)^2\right\rangle = [\partial_\mathrm{z} C_\mathrm{m}\cdot V_\mathrm{g}^\mathrm{DC} V_\mathrm{g}^\mathrm{AC} Q_\mathrm{m}/(m_\mathrm{eff}\omega_\mathrm{m}^2)]^2$.
Remarkably, $g_0$ gets larger upon increasing $|V_\mathrm{g}^\mathrm{DC}|$ for device $A$. This tunability of $g_0$ is attributed to the static deflection of the graphene sheet. 
When incorporating the effect of the static displacement into $C_\mathrm{m}$, we get a good agreement between the expected $g_0 = \omega_\mathrm{c}/(2C)\cdot \partial_\mathrm{z}C_\mathrm{m}$ and the measurements, using $C = 75$~fF and 100~fF for devices $A$ and $B$, respectively (red lines in Fig.~3f,g). These values of $C$ agree well with $C = 90~$fF estimated from simulations.
The obtained coupling rates $g_0$ compare favourably with previous experiments carried out with mechanical resonators made from other materials. Indeed, the coupling was $g_0/2\pi \sim 1~$Hz in works with cavity geometries similar to ours~\cite{teufel08,roch10,zhou2013}. Larger values were achieved with closed-loop cavities 
($g_0/2\pi =40\, \mathrm{and}\,210~$Hz) but this geometry does not allow one to apply $V_\mathrm{g}^\mathrm{DC}$ between the mechanical resonator and a counter electrode as discussed above~\cite{teufel2011b,massel2011}.

\begin{figure}[htb]
\centering
\includegraphics[width=\linewidth]{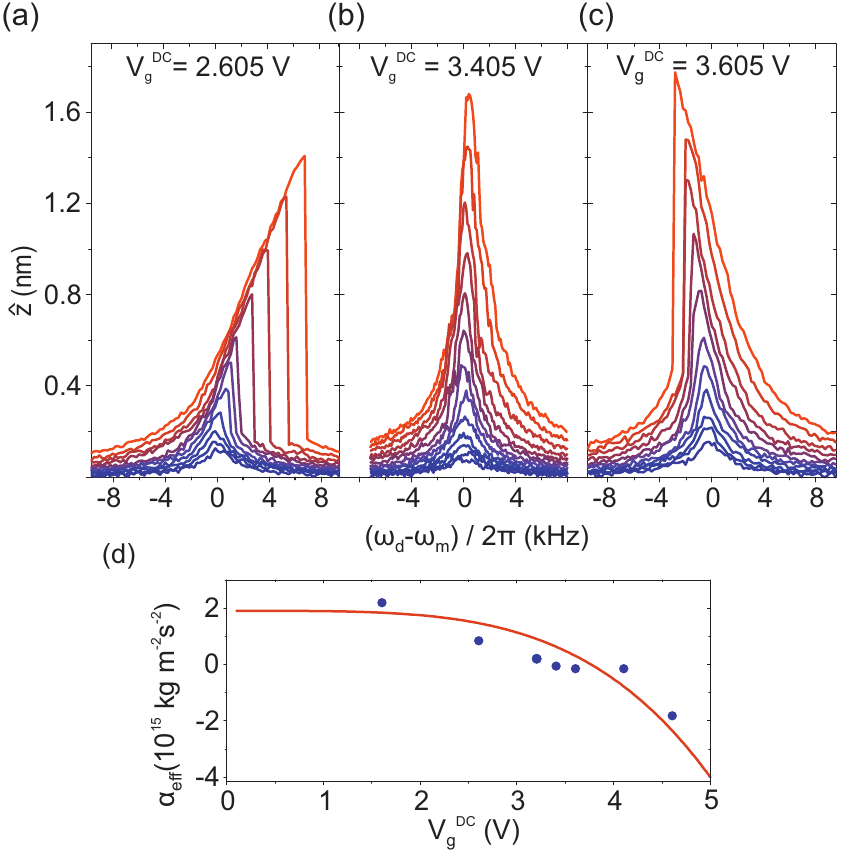}
\caption[Figure 5] {
(a)-(c) Dependence of the vibrational amplitude $\hat{z}$ in device $B$ on the drive frequency for different $V_\mathrm{g}^\mathrm{AC}$ in each plots. In (a) and (c), $V_\mathrm{g}^\mathrm{AC} = 3~\mu$V - 31~$\mu$V. In (b) $V_\mathrm{g}^\mathrm{AC} = 1.9\mu V - 31\mu V$. The onset of bistability is determined to be at $\hat{z}_\mathrm{crit} = 360~$pm for (a) and at $\hat{z}_{\mathrm{crit}} = 900~$pm for (c). (d) Effective Duffing parameter $\alpha_{\mathrm{eff}}$ as a function of $V_\mathrm{g}^\mathrm{DC}$. The red line is a plot of Eq.~\eqref{eq:alphaeff} with $c_\mathrm{s} = 0.65$~nm/V$^2$ and $\alpha_0 = 1.9 \cdot 10^{15}~$kg$\cdot$m$^{-2}$s$^{-2}$.
}
\label{f:figure 5}
\end{figure}

Now, we investigate how the strong tunability of the graphene equilibrium position affects the nonlinear response of the mechanical resonator. 
For this, we measure $P_\mathrm{out}$ as a function of $\omega_\mathrm{d}$ as in Fig.~3b,c in order to obtain the response of the vibrational amplitude $\hat{z}(\omega_\mathrm{d})$ for large driving forces at different $V_\mathrm{g}^\mathrm{DC}$ (Fig.~5a-c). Interestingly, we are able to tune the sign of the Duffing nonlinearity from a hardening behavior at low $V_\mathrm{g}^\mathrm{DC}$ (Fig.~5a) to a softening behavior at high $V_\mathrm{g}^\mathrm{DC}$ (Fig.~5c). At an intermediate $V_\mathrm{g}^\mathrm{DC}$ of  about $3.4$~V, we are able to cancel the Duffing nonlinearity, that is, the resonant frequency remains roughly constant upon varying the driving force (Fig.~5b).
We quantify the Duffing nonlinearity from the critical displacement amplitude $\hat{z}_\mathrm{crit}$ above which the response gets bistable. 
For a Duffing resonator with linear damping, 
the effective Duffing constant $\alpha_\mathrm{eff}$ is related to $\hat{z}_\mathrm{crit}$ by
$\alpha_\mathrm{eff} = \frac{8}{3\sqrt{3}}m_\mathrm{eff}\omega_\mathrm{m}^2/(Q_\mathrm{m} \hat{z}_\mathrm{crit})$~\cite{lif08}.
Figure~5d shows that $\alpha_\mathrm{eff}$ is positive at low $V_\mathrm{g}^\mathrm{DC}$ and becomes negative at large $V_\mathrm{g}^\mathrm{DC}$.
This dependence can be attributed to the symmetry breaking of the mechanical motion induced by static deflection~\cite{younis03,eichler2013symmetry}, which reads
\begin{equation}
\alpha_\mathrm{eff} \approx \alpha_0 -\frac{10}{m_\mathrm{eff}\omega_\mathrm{m}^2}\alpha_0^2 z_\mathrm{s}^2 
\label{eq:alphaeff}
\end{equation}
where $\alpha_0$ is the Duffing constant when $V_\mathrm{g}^\mathrm{DC} = 0$; $\alpha_0$ could have a geometrical origin~\cite{lif08}. The fit of Eq.~\eqref{eq:alphaeff} to the measurement yields $c_\mathrm{s} = 0.65$~nm/V$^2$ and $\alpha_0 = 1.9 \cdot 10^{15}~$kgm$^{-2}$s$^{-2}$ (red line in Fig.~5d). This value of $c_\mathrm{s}$ is consistent with that expected from Eq.~\eqref{eq:staticdisp}.
The sign change of the Duffing nonlinearity due to static deformation is a unique property of graphene and nanotube resonators~\cite{eichler12}.

The prospects to reach the quantum regime with graphene resonators are promising.
For this, it is illustrative to compare the figures of merit achieved here to those reported by Teufel~\textit{et al.}~\cite{teufel2011b}, which demonstrated ground-state cooling with a superconducting cavity. 
In the device $A$ of our work, we measure $g_0/2\pi \approx 15~$Hz, $n_\mathrm{p} = 8000$, $Q_\mathrm{m} = 100,000$ and $\kappa_\mathrm{int}/2\pi = 13~$MHz, while the parameters of Teufel~\textit{et al.} are $g_0/2\pi \approx 200~$Hz, $n_\mathrm{p} = 4000$, $Q_\mathrm{m} = 350,000$ and $\kappa_\mathrm{int}/2\pi = 40~$kHz. As discussed above, an obvious way to improve $\kappa_{int}$ is to fabricate cavities with less contamination and imperfections. $\kappa_{int}$ can then be further reduced by lowering the resistance of the graphene flake. This can be achieved for instance by selecting thicker graphene flakes or electrostatically doping the graphene. Minimizing the graphene resistance, together with increasing the area of the interface between the graphene and the electrodes, is beneficial for diminishing Joule heating at high pump power. In order to increase $g_0$, we will reduce $d$ further by fabrication and graphene pulling. We should reach $g_0/2\pi \approx 250~$Hz with $d = 30$~nm. An alternative route to increase $g_0$ is to enhance the coupling using a cooper-pair box~\cite{heikkila2013,rimberg2013}.

In conclusion, we have reported devices where a graphene resonator is coupled to a superconducting cavity. 
The tunability of these devices, in combination with the large graphene-cavity coupling, constitutes a promising approach to study quantum motion.
The large reduction of the resonance frequency of the graphene resonator observed here is interesting to enhance the zero-point motion and to increase the effect of mechanical nonlinearities~\cite{voje12,rips13,rips14}. 
The tunability of the resonance frequency with $V_\mathrm{g}^\mathrm{DC}$ is suitable for parametric amplification and quantum squeezing of mechanical states~\cite{xin2014}. 
In these graphene-cavity devices, the opto-mechanical coupling can be varied not only with the number of pump photons but also with $V_\mathrm{g}^\mathrm{DC}$. Interestingly, the tuning of the coupling with $V_\mathrm{g}^\mathrm{DC}$ can be made faster than that with $n_\mathrm{p}$, since the inverse of the cavity linewidth poses an upper limit on how fast the photon number inside the cavity can be changed. Because the mass of graphene is ultra-low, its motion is extremely sensitive to changes in the environment. It will be interesting to couple the quantum vibrations of motion to other degrees of freedom, such as electrons and spins.\\

\textbf{Author contributions}\\
PW, JG and IT developed the fabrication process.
PW fabricated the devices with support from JG.
The experimental setup was built up by JG with support from PW.
PW and JG carried out the measurements.
JG and PW analyzed the data with support from AB and DEC.
JG and AB wrote the manuscript with critical comments from all authors.
AB and JG conceived the experiment and supervised the work. \\

\textbf{Acknowledgements}\\
We would like to thank Joel Moser, Gabriel Puebla, Christopher Eichler, Andreas Isacsson, Martin Eriksson, Sara Hellm\"uller and Andreas Wallraff for helpful discussions.
We gratefully acknowledge Gustavo Ceballos and the ICFO mechanical and electronic workshop for support. 
We acknowledge support from the European Union through the RODIN-FP7 project, the ERC-carbonNEMS project, and the Graphene Flagship (grant agreement 604391), the Spanish state (MAT2012-31338), the Catalan government (AGAUR, SGR).

\newpage

\bibliography{referencesCavityGraphene}

\newpage
\widetext
\begin{center}
\textbf{\large Supplementary material to: Coupling graphene mechanical resonators to superconducting microwave cavities}

\vspace{5pt}
\large P. Weber, J. G\"uttinger, I. Tsioutsios, D.E. Chang, A. Bachtold
\vspace{5pt}

\textit{\large ICFO-Institut de Ciencies Fotoniques,Mediterranean Technology Park, 08860 Castelldefels (Barcelona), Spain}
\end{center}
\setcounter{equation}{0}
\setcounter{figure}{0}
\setcounter{table}{0}
\setcounter{page}{1}
\makeatletter
\renewcommand{\theequation}{S\arabic{equation}}
\renewcommand{\thefigure}{S\arabic{figure}}
\renewcommand{\bibnumfmt}[1]{[S#1]}
\renewcommand{\citenumfont}[1]{S#1}

%

\section{Fabrication of the superconducting structure}
We use a highly resistive silicon substrate (6 k$\Omega$cm) with a 295~nm thick, dry chlorinated thermal oxide from \emph{NOVA wafers}.
The wafers are sputtered with 200~nm Nb, followed by optical lithography and ion-milling to define the superconducting cavity, the feedline and the graphene contacts. These process steps, and the subsequent wafer dicing, are carried out by \emph{STAR cryoelectronics}.
We use Nb as a cavity material because of the high critical temperature $T_c = 9.2$~K that allows the cavity to be tested at liquid helium temperature and to sustain large pump fields. 
The fine structure of the device, shown in Fig.~1a of the letter, consists of the cavity counter electrode and the support electrodes used later on to anchor the graphene flake. The fabrication of this fine structure is carried out with electron-beam lithography (EBL) and reactive-ion etching (RIE).
In a first EBL/RIE step, the cavity counter electrode is separated from the support electrodes. As a mask for etching, we use 50~nm aluminium (Al).
The Al-mask is structured with EBL using PMMA and etched in 0.2\% Tetra-Methyl-Ammonium-Hydroxide (TMAH) diluted in H$_2$O. Unmasked areas are cleaned from Al-residues with 30~s ion-milling in an argon (Ar) atmosphere. The Nb is etched with RIE in a 10~mTorr SF$_6$/Ar atmosphere with a radio frequency (RF) power of 100~W. In a second EBL/RIE step the cavity counter electrode is thinned down, such that the height difference between the cavity counter electrode and the support electrodes equals $d$.

Here we would like to comment as well on the gap between the two support electrodes, which contact the graphene (Fig.~1a). On the one hand this gap allows measuring electrical transport through the graphene, on the other hand it helps in preventing the collapse of the graphene against the cavity counter electrode during critical point drying. The openings did not show a significant influence on the mechanical behaviour in numeric simulations (private communication with Andreas Isacsson and Martin Eriksson).

\section{Characterization of the electrical setup and the cavity}
\subsection{Calibration of loss and gain in the input and output lines of the cryostat}
To relate the externally applied RF power and the measured RF power to the actual fields at the sample, a careful calibration of the attenuation and gain in the setup is needed. The RF-input lines are attenuated at different temperature stages in the cryostat to shield the device from electromagnetic noise and to thermalize the lines.
The attenuation is 10~dB at $T = 47~$K, 20~dB at $T = 4~$K, 6~dB at $T = 700~$mK and 20~dB at $T = 30~$mK, where we use for the last attenuation step a directional coupler to physically interrupt the central part of the coaxial line~\cite{Teufel2011bSI}. 
The total loss in the lines is the sum of the contributions from the attenuators and the loss in cables and connectors.
In the input lines of the cold cryostat we measure a total attenuation of $loss(\omega_\mathrm{d}) = 57$~dB in the 10-100~MHz range and $loss(\omega_\mathrm{c}) = 64$~dB around $\omega_{\mathrm{c}}/2\pi = 6.7~$GHz.
The output of the cavity is shielded by two \emph{QUINSTAR CTH0408KC} circulators that are operated as isolators at 30~mK, and then amplified by a low-noise amplifier \emph{LNF-LNC4\_8A} from \emph{Low Noise Factory} at 4~K with gain $G(\omega_c) = 43~$dB and noise temperature $T_\mathrm{noise} \approx 2$~K measured by the factory at 10~K. 

We measure a detection limit of $S_{N,SA} = -157$~dBm/Hz in our spectrum analyzer (SA). This noise floor is limited by the input noise of the amplifier. From $k_\mathrm{B} T_\mathrm{noise}(G-loss_\mathrm{4K-SA}) = -157$~dBm we can extract $G-loss_\mathrm{4K-SA}  = 38.5$~dB and $loss_\mathrm{4K-SA} = 4.5$~dB. The total measured gain of the output line is $gain =G-loss_\mathrm{4K-SA}-loss_\mathrm{sample-4K} \approx 35~$dB with $loss_\mathrm{sample-4K} \approx 3.5~$dB which is reasonable considering the losses in the two circulators and the line at the level of the sample. Hence we are able to resolve noise powers of $S_\mathrm{N} >-157~$dBm/Hz$-35~$dB$ = -192~$dBm/Hz in the transmission line of the sample. 
We measure a total transmitted power of $| S_{\mathrm{21}} |^2 = -29.5~$dB in device $A$ and $| S_{\mathrm{21}} |^2 = -28.5~$dB in device $B$, from the output of the RF source to the input of the spectrum analyzer around $\omega_\mathrm{c}$. This values agree well with our calibration, $| S_{\mathrm{21}} |^2 = -loss+gain = -29~$dB. 

\subsection{Analyzing the cavity lineshape}
To characterize the external coupling and the internal loss of the cavity, we measure the lineshape of the resonance of the cavity.
The normalized transmission is given by~\cite{mazin2005SI}
\begin{equation}
S_{\mathrm{21}}(\delta \omega_\mathrm{c}) = 1 - \frac{\kappa_\mathrm{ext}/\kappa}{1+2i \delta \omega_\mathrm{c}/\kappa}.
\label{eqref:S21}
\end{equation}
with $\delta \omega_\mathrm{c}$ the detuning from the cavity resonance frequency $\omega_\mathrm{c}$, $\kappa$ the total cavity decay rate and $\kappa_\mathrm{ext}$ the external coupling rate of the cavity to the feedline. At resonance ($\delta \omega_{\mathrm{c}} = 0$) the transmission is 
\begin{equation}
S_{\mathrm{21,min}} = - \frac{\kappa_\mathrm{ext}}{\kappa_\mathrm{int}},
\end{equation}
with $\kappa_\mathrm{int}$ the internal cavity decay due to cavity internal losses.
By measuring the depth and the width of the transmission dip, $\kappa$, $\kappa_\mathrm{ext}$ and $\kappa_\mathrm{int}$ are extracted.
\begin{figure*}[hbt]
\includegraphics[width=0.45\textwidth]{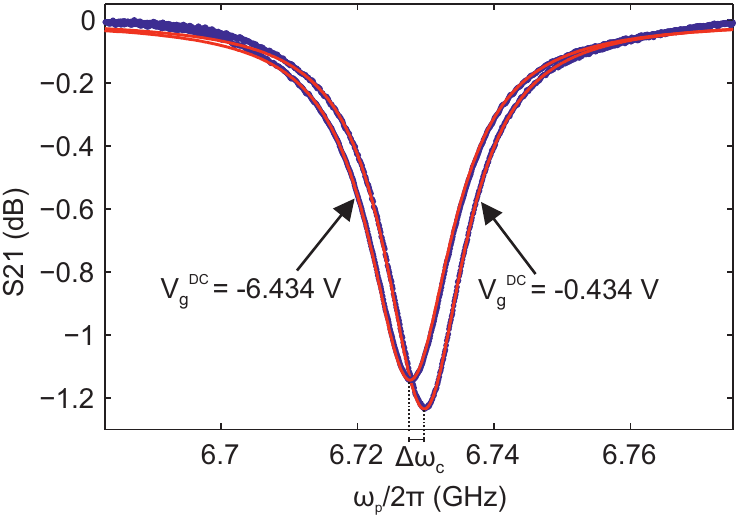}
\caption{Transmission through the feedline around the cavity resonance frequency.}
\label{f:figureS21}
\end{figure*} 
In Fig. S1 we show the measured transmission spectrum of device $A$. For this plot we subtract the background of the measurement containing contributions from the input and the output lines. From a fit of the spectrum to Eq.~\eqref{eqref:S21} we extract for $V_\mathrm{g}^\mathrm{DC}=-0.434~$V the resonance frequency of the cavity $\omega_\mathrm{c}/2\pi=6.73~$GHz and the external and internal decay rates $\kappa_\mathrm{ext}=2~$MHz and $\kappa_\mathrm{int}=13.2~$MHz. By changing $V_\mathrm{g}^\mathrm{DC}$ to $-6.434~$V we observe a decrease in the resonance frequency $\Delta \omega_\mathrm{c}/2\pi \approx 2~$MHz, which is equivalent to a change in cavity capacitance of $\Delta C \approx 50~$aF. In addition, the internal decay rate of the cavity increases to $\kappa_\mathrm{int}=14~$ MHz. The change in resonance frequency can be well explained with an increased graphene-cavity capacitance due to static displacement
\begin{equation}
\Delta C_\mathrm{m} = \int_\mathrm{0}^{2\pi} d\phi \int_\mathrm{0}^{R_\mathrm{g}}r dr {\frac{\epsilon_\mathrm{0}}{d-\xi_\mathrm{s}(r)}} -C_\mathrm{m0}, 
\end{equation}
where $C_\mathrm{m0}$ is the capacitance of the graphene without static displacement at $V_\mathrm{g}^\mathrm{DC} \approx 0~$V and $\xi_\mathrm{s}(r)=z_\mathrm{s}(V_\mathrm{g}^\mathrm{DC}) \cdot (r^2/R^2-1)$ (see S.4.2) the static mode shape of the pulled down graphene with $z_\mathrm{s}(V_\mathrm{g}^\mathrm{DC})$ the deflection of the center point of the membrane. If we calculate the capacitive change using $z_\mathrm{s}(V_\mathrm{g}^\mathrm{DC})=15~$nm for $V_\mathrm{g}^\mathrm{DC} \approx 6~$V (Fig. 4a, main text) we obtain $\Delta C_\mathrm{m}=49~$aF. This value is in excellent agreement with the value of $\Delta C$ estimated from the change in $\omega_\mathrm{c}$. 

\subsection{Modeling the dissipation of the cavity}

\begin{figure*}[hbt]
\includegraphics[width=1\textwidth]{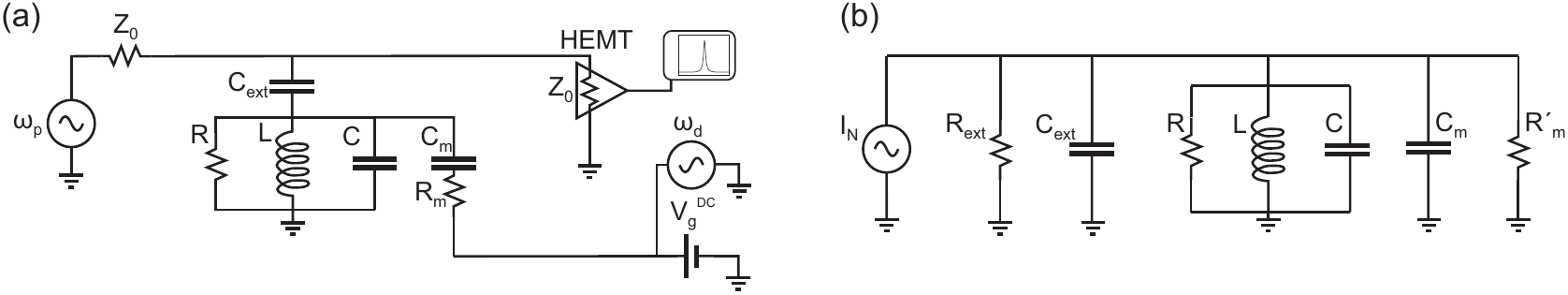}
\caption{(a) Equivalent circuit of the measurement scheme. (b) Norton equivalent circuit where all the contributions are converted into a parallel RLC circuit. }
\label{f:figureS2}
\end{figure*}

In Fig.~S2a we show a detailed equivalent circuit of our measurement setup. In order to model dissipation we included (i) an input and output impedance of $Z_0=50~\Omega$ in the RF source and the cryogenic amplifier, (ii) a resistor $R_\mathrm{m}$ for the loss in the graphene and the DC connection and (iii) a resistor $R$ for the internal loss in the cavity. By using a Norton equivalent circuit~\cite{goeppl2008SI} we can convert all contributions into a parallel equivalent RLC circuit (Fig.~S2b) with $1/R_\mathrm{tot}=1/R_\mathrm{ext}+1/R+1/R_\mathrm{m}'$ and $C_\mathrm{tot}=C_\mathrm{ext}+C+C_\mathrm{m}$. We obtain $1/R_\mathrm{ext} \approx \omega^2C_\mathrm{ext}^2 Z_0/2$, $I_\mathrm{N} = V_\mathrm{p}i\omega C_\mathrm{ext}$, $1/R_\mathrm{m}' \approx \omega^2 C_\mathrm{g}^2 R_\mathrm{m}$ and we have made use of the fact that in our circuit $\omega C_\mathrm{ext} Z_0 \ll 1$ and $\omega^2R_\mathrm{m}^2C_\mathrm{g}^2 \ll 1$. The linewidth of the equivalent parallel RLC-circuit is then given by
\begin{equation}
\kappa 
= \frac{1}{C_\mathrm{tot}R_\mathrm{tot}} = \underbrace{\frac{1}{C_\mathrm{tot}R_\mathrm{ext}}}_{\kappa_{ext}}+\underbrace{\frac{1}{C_\mathrm{tot}R}}_{\kappa_\mathrm{cavity}}+\underbrace{\frac{1}{C_\mathrm{tot}R_\mathrm{m}'}}_{\kappa_\mathrm{g}}
\label{eqref:kappa}
\end{equation}
By substituting the equivalent resistances we get
\[
\kappa_{ext} = \frac{\omega_c^2C_\mathrm{ext}^2 Z_0}{2C_{tot}} \qquad \mathrm{and} \qquad \kappa_g = \frac{\omega_c^2C_g^2R_m}{C_{tot}}.
\]
From the measured external linewidth we estimate the coupling capacitance using the above expression to be
\[
C_\mathrm{ext} = \sqrt{\frac{2 C_\mathrm{tot} \kappa_\mathrm{ext}}{\omega_\mathrm{c}^2 Z_0}}.
\]
Using $C_\mathrm{tot} = 90$~fF, $\kappa_\mathrm{ext}/2\pi = 2~$MHz, $\omega_\mathrm{c} = 6.7$~GHz and $Z_0 = 50$~$\Omega$ we get $C_\mathrm{ext} = 5$~fF in good agreement with the simulated values of $C_\mathrm{ext} = 4$~fF for device $A$ and $C_\mathrm{ext} = 6$~fF for device $B$.

Furthermore, the measured increase of $\kappa$ with $V_\mathrm{g}^\mathrm{DC}$ in Fig.~S1 allows to estimate the resistance $R_m$ in device A. If we assume that the whole change of the linewidth is due to the static displacement of the resonator ($\Delta \kappa= \Delta \kappa_\mathrm{g}$) we have
\[ 
R_\mathrm{m} = \frac{\Delta \kappa_\mathrm{g} C_\mathrm{tot}}{\omega_\mathrm{c}^2  (C_\mathrm{m}^2-C_\mathrm{m0}^2)}.
\]
Using $\Delta \kappa = \Delta \kappa_\mathrm{g}$, $\Delta \kappa_\mathrm{g}/2\pi = 0.8~$MHz,  $C_\mathrm{m} = 0.4~$fF and $C_\mathrm{m0} = 0.35~$fF (from section S.2.2) we obtain $R_\mathrm{m} \approx 6~$k$\Omega$. Here, the change in the capacitance $C_\mathrm{m}$ is derived from the measured change of the cavity resonance frequency $\Delta \omega_c$. By inserting the value for $R_\mathrm{m}$ in equation \eqref{eqref:kappa} we obtain $\kappa_\mathrm{cavity}/2\pi=10.8~$MHz and $\kappa_\mathrm{g}/2\pi=2.4~$MHz with $\kappa_\mathrm{int}=\kappa_\mathrm{cavity}+\kappa_\mathrm{g}$. The high value of $\kappa_\mathrm{int}$ is therefore mainly attributed to the contamination and imperfections of the cavities. Indeed, we have tested the cavity of devices A and B at $T=4.2~$K before the transfer of the graphene flakes, and we observed larger $\kappa_\mathrm{int}$ than what we usually observe in devices processed in the same way.

\section{Coupling of the graphene resonator to the superconducting cavity}
\subsection{Estimation of the coupling parameter $g_0$} \color{black}
By applying a pump power $P_\mathrm{p,in}$ at frequency $\omega_\mathrm{p}$ at the feedline, we create a cavity photon population of 
\[
n_\mathrm{p} = \frac{1}{\hbar\omega_\mathrm{p}}P_\mathrm{p,in} \cdot \frac{2}{\kappa_\mathrm{ext}} \cdot \frac{\kappa_\mathrm{ext}^2}{\kappa^2+4(\omega_\mathrm{p}-\omega_\mathrm{c})^2}.
\]
Here, $\kappa_\mathrm{ext}/2$ is the coupling of the input mode of the feedline to the cavity and $\kappa_\mathrm{ext}^2/(\kappa^2+4(\omega_\mathrm{p}-\omega_\mathrm{c})^2)$ is the lineshape of the cavity resonance. 
The photon population in the cavity interacts by Stokes (-) and anti-Stokes (+) scattering with the mechanical resonator. The scattering rates are given by
\[
\Gamma_{\pm} = 4 n_\mathrm{p} g_0^2\frac{\kappa}{\kappa^2+4(\omega_\mathrm{p}-\omega_\mathrm{c} \pm \omega_\mathrm{m})^2}, 
\]
with $g_0$ the single-photon coupling and $\omega_\mathrm{m}/2\pi$ the mechanical resonance frequency. 
In the case of $\omega_\mathrm{p}-\omega_\mathrm{c} = - \omega_m$ anti-Stokes scattering is resonantly enhanced and we have $\Gamma_\mathrm{opt} \approx 4 n_\mathrm{p} g_0^2/\kappa$ in the so-called resolved sideband limit where $\omega_\mathrm{m} \gg \kappa$. Here we introduced $\Gamma_\mathrm{opt}$, the opto-mechanical coupling rate.

The anti-Stokes scattering leads to an equilibrium cavity population $n_\mathrm{c}$ at $\omega_\mathrm{c}$ determined by $n_c\kappa \approx \Gamma_\mathrm{opt}n_\mathrm{m}$. Here, we assumed negligible thermal population of the superconducting cavity (at 30~mK $n_\mathrm{c,th} = 1/(\exp{\hbar\omega_\mathrm{c}/k_B T}-1) \ll 1$) and negligible direct population due to the phase noise of the pump. The number of phonons $n_\mathrm{m}$ is related to the zero-point motion $z_\mathrm{zp}$ by $n_\mathrm{m} \approx \left\langle z\right\rangle^2/2z_\mathrm{zp}^2$, where $\left\langle z\right\rangle$ is the time averaged deflection of the effective mass motion. The cavity mode leaks into the output mode of the feedline with a rate $\kappa_\mathrm{ext}/2$ and results in the detectable output power $P_\mathrm{out} = n_\mathrm{c} \hbar\omega_\mathrm{c} \kappa_\mathrm{ext}/2$ or
\[
P_\mathrm{out}(\omega_\mathrm{c}) = P_\mathrm{p,in} \cdot \frac{\kappa_\mathrm{ext}^2}{\kappa^2+4(\omega_\mathrm{p}-\omega_\mathrm{c})^2}\cdot\left(\frac{1}{\kappa}\frac{\partial\omega_\mathrm{c}}{\partial x}\right)^2\cdot 2\left\langle z^2\right\rangle.
\]
From the measured output power we can then estimate $g_0$ as
\begin{equation}
g_0 = z_\mathrm{zp}\sqrt{P_{out}(\omega_\mathrm{c})\frac{\kappa^2}{n_\mathrm{p}\hbar\omega_\mathrm{c}\kappa_\mathrm{ext}\left\langle z^2\right\rangle }}. 
\end{equation}

To model the dependence of $g_0 = G_0 z_\mathrm{zp}$ on the voltage $V_\mathrm{g}^\mathrm{DC}$ between the graphene and the cavity counter electrode, we have to account for the $V_\mathrm{g}^\mathrm{DC}$ dependence of both $G_0$ and $z_\mathrm{zp}$. To estimate $G_0(V_\mathrm{g}^\mathrm{DC})$ we use the calculated value of the equilibrium position $z_\mathrm{s}$ (Fig.~4a of the letter) to substitute $d$ by $d = d_0-z_\mathrm{s}$ in the calculated graphene-cavity capacitance $C_\mathrm{m}$
\[
G_0(V_\mathrm{g}^\mathrm{DC}) = \frac{\omega_{c}}{2 C_\mathrm{tot}} \frac{\partial C_\mathrm{m} (V_\mathrm{g}^\mathrm{DC})}{\partial z} \approx \frac{\omega_c}{2C}\frac{0.433\pi R_\mathrm{g}^2 }{[d_0-z_\mathrm{s}(V_\mathrm{g}^\mathrm{DC})]^2},
\]
where $d_0$ is the value of $d$ at $V_\mathrm{g}^\mathrm{DC}=0$, $C_\mathrm{tot}$ is the total cavity capacitance approximated by the cavity capacitance $C$ and $R_\mathrm{g}$ is the radius of the cavity counter electrode. The factor 0.433 is a correction due to the effective mass modeling (see section S.4.4).
For the calculation of the capacitance $C_\mathrm{m}$ see below the section about the effective mass modelling.
The increase of the zero-point motion is accounted for by calculating $z_\mathrm{zp}$ as a function of $V_\mathrm{g}^\mathrm{DC}$ from the measurement of the resonance frequency $\omega_\mathrm{m}$ as a function of $V_\mathrm{g}^\mathrm{DC}$ in Fig.~3c,d of the letter
\[
z_\mathrm{zp}(V_\mathrm{g}^\mathrm{DC}) = \sqrt{\frac{\hbar}{2 m_\mathrm{eff} \omega_\mathrm{m}(V_\mathrm{g}^\mathrm{DC})}}.
\]

\subsection{Displacement sensitivity}
The detection limit of our readout circuit $S_\mathrm{N} = -192~$dBm/Hz (see section S.2.1) imposes a limit on the measurement imprecision $\sqrt{S_\mathrm{z,imp}}$ with
\[
S_\mathrm{z,imp} = \frac{S_\mathrm{N} \kappa^2 z_\mathrm{zp}^2}{n_p \hbar\omega_\mathrm{c}\kappa_\mathrm{ext}g_0^2}.
\]
For the parameters in device $A$ we get $\sqrt{S_\mathrm{z,imp}} = 2.55~\mathrm{pm}/\sqrt{\mathrm{Hz}}$ at $n_\mathrm{p} = 8000$ and $\sqrt{S_\mathrm{z,imp}} = 230~\mathrm{fm}/\sqrt{\mathrm{Hz}}$ at $n_\mathrm{p} = 10^6$. For comparison, the height of the resonance in the power spectral density of the thermal motion at 1~K is $(42~\mathrm{fm}/\sqrt{\mathrm{Hz}})^2$ ($(7~\mathrm{fm}/\sqrt{\mathrm{Hz}})^2$ at 30~mK). We will improve our displacement resolution (i) by reducing the loss in the cavity (up to a factor 8 improvement in $\sqrt{S_\mathrm{z,imp}}$), (ii) by using a quantum limited amplifier~\cite{castellanos2008SI} (up to a factor 10 improvement in $\sqrt{S_\mathrm{z,imp}}$) and (iii) by increasing the coupling (with a factor 10 improvement in $\sqrt{S_\mathrm{z,imp}}$ for $g_0/2\pi = 70$~Hz).

\section{Mechanical modelling}
\subsection{Circular graphene resonator in the membrane limit}
We model the deflection $\xi(t,x,y)$ of the graphene resonator as a thin plate subject to large external stretching (membrane limit)~\cite{landau1970SI}
\begin{equation}
\rho_\mathrm{2D} \frac{\partial^2 \xi}{\partial t^2}=T \nabla^2 \xi+P(x,y),
\end{equation}
with $\rho_\mathrm{2D}$ the sheet mass density, $P(x,y)$ the local pressure in z-direction and $T$ a stretching force per unit length at the edge of the membrane.
If we consider radially symmetric modes $\xi(t,r)$, the stretching force $T$ is related to a radial strain $\epsilon = (R'-R)/R$ with the elongated radius $R'$ by 
\begin{equation}
T = Eh \epsilon = Et n_\mathrm{g}  \epsilon, 
\end{equation}
with the Young's modulus of graphite $E \approx 1$~TPa or the two dimensional graphene Young's modulus $Et = 340~$N/m ~\cite{lee2008SI}, $n_\mathrm{g}$ the number of graphene layers and $t = 0.335$~nm~\cite{jishi82SI} the interlayer spacing in graphite.
The total sheet mass density $\rho_\mathrm{2D} = \eta n_\mathrm{g} \rho_\mathrm{graphene}$ includes the mass from the graphene layers, with the graphene mass density $\rho_\mathrm{graphene} = 7.6\times10^{-19}~$kg$/\mu$m$^2$, and a correction factor $\eta \geq 1$ to account for additional adsorbents on the graphene.

The electrostatic pressure due to the gate voltage is modelled in a parallel plate approximation with the capacitive energy given by $U=\frac{1}{2}C_\mathrm{m} V_\mathrm{g}^2$. If we expand the capacitance in terms of $\xi$ we get 
\begin{eqnarray*}
	U &\approx & \int {dx dy \frac{\epsilon_0 V_\mathrm{g}^2}{2}\frac{1}{d-\xi(x,y)}}\\
		&\approx & \int {dx dy \frac{\epsilon_0 V_\mathrm{g}^2}{2d}\left( 1 +\frac{\xi(x,y)}{d}+\frac{\xi(x,y)^2}{d^2}+\frac{\xi(x,y)^3}{d^3}+\ldots \right)}\\
	\frac{\partial U}{\partial z}&\approx & \int {dx dy \frac{\epsilon_0 V_\mathrm{g}^2}{2d^2}\left(1 +\frac{2\xi(x,y)}{d}+\frac{3\xi(x,y)^2}{d^2}+\frac{4\xi(x,y)^3}{d^3}+\ldots \right)}\\
	&\approx & \int {dx dy P(x,y)}.
\end{eqnarray*}
The differential equation for the deflection is then given by
\begin{equation}
\rho_\mathrm{2D}\frac{\partial^2 \xi}{\partial t^2} =  T\nabla^2 \xi - \frac{\epsilon_0 V_\mathrm{g}^2}{2d^2}\left(1 +\frac{2\xi(x,y)}{d}+\frac{3\xi(x,y)^2}{d^2}+\frac{4\xi(x,y)^3}{d^3}+\ldots \right)
\label{eq:localmotion}
\end{equation}

To solve the equation, we decompose the deflection $\xi(r,t)$ into a static displacement $\xi_\mathrm{s}(r)$ and time-dependent (radial) modes $k$ with amplitude $\xi_\mathrm{k}(r)$ 
\[
\xi(r,t) \approx \xi_\mathrm{s}(r)+\sum_k{\xi_\mathrm{k}(r)e^{-i\omega t}}.
\]

\subsection{Static displacement as a function of DC voltage} 
For the static displacement we have 
\[
0 = T \nabla^2 \xi_\mathrm{s}(r) - \frac{\epsilon_0 (V_\mathrm{g}^\mathrm{DC})^2}{2d^2}\left(1+\frac{2\xi_\mathrm{s}(r)}{d} + \dots \right)
\] 
by assuming $2\xi_\mathrm{s}(r)/d \ll 1$. The solution at lowest order in $\xi_\mathrm{s}(r)/d$ is given by
\[
\xi_\mathrm{s} (r) = \frac{\epsilon_0 (V_\mathrm{g}^\mathrm{DC})^2}{8T d^2}\left(r^2-R^2\right)
\]
with the normalized center deflection 
\begin{equation}
z_\mathrm{s} = \frac{\epsilon_0 R_\mathrm{g}^2}{8T d^2}(V_\mathrm{g}^\mathrm{DC})^2 = c_\mathrm{s} (V_\mathrm{g}^\mathrm{DC})^2.
\label{eq:staticdispSI}
\end{equation}
For device $A$, the approximation of small static deflections is well valid up to $V_\mathrm{g}^\mathrm{DC} \approx 3.5$~V where $z_\mathrm{s} = 5$~nm and $2z_\mathrm{s}/d = 0.1 \ll 1$. At large $V_\mathrm{g}^\mathrm{DC}$ we underestimate the static displacement by not including higher order corrections of the electric force. On the other hand we also underestimate the mechanical force when neglecting nonlinear effects as described below. In device $B$ $z_\mathrm{s} \approx 7.5$~nm with $2z_\mathrm{s}/d = 0.1 \ll 1$, which corresponds to $V_\mathrm{g}^\mathrm{DC} \approx 5$~V.

The assumption of constant strain at moderate gate voltages is justified by analysing the strain induced by the static deflection. At $V_\mathrm{g}^\mathrm{DC} \approx 6$~V the additional strain induced by the static deflection of 10~nm (in device $B$) is given by $\epsilon_\mathrm{s} = 2\cdot 10^{-5} \ll \epsilon_\mathrm{init}$, significantly smaller than the initial strain.

\subsection{Mechanical resonance frequency as a function of gate voltage}
If we assume orthogonal modes and neglect mode coupling, we can project Eq.~(\ref{eq:localmotion}) on the fundamental mode
\begin{equation}
-\rho_\mathrm{2D}\omega^2 \xi_\mathrm{f}(r) = T\nabla^2 \xi_\mathrm{f}(r) - \frac{\epsilon_0 (V_\mathrm{g}^\mathrm{DC})^2}{d^3}\xi_\mathrm{f}(r)
\end{equation}
and solve for the mode amplitude $\xi_\mathrm{f}(r)$.
Considering a clamped boundary with $\xi_\mathrm{f}(R) = 0$ we get
\begin{equation}
\xi_\mathrm{f}(r) = \hat{z} J_0\left(\frac{2.4}{R} r\right),
\end{equation}
where $\hat{z} = \xi_\mathrm{f}(0)$ is the deflection amplitude at the center of the membrane and $J_0$ is the 0th Bessel function with $J_0(2.4) = 0$. The resonance frequency as a function of gate voltage is then given by
\begin{equation}
\omega_\mathrm{m}(V_\mathrm{g}^\mathrm{DC}) = \sqrt{\frac{2.4^2 T}{R^2\rho_\mathrm{2D}} - \frac{\epsilon_0}{d^3\rho_\mathrm{2D}} (V_\mathrm{g}^\mathrm{DC})^2}.
\end{equation}
By taking into account the reduced radius of the cavity counter electrode $R_\mathrm{g}$ with respect to the membrane radius $R$, the electrical force gets reduced by a factor $R_\mathrm{g}^2/R^2$ and we obtain
\begin{equation}
\omega_\mathrm{m}(V_\mathrm{g}^\mathrm{DC}) = \sqrt{\frac{2.4^2T}{R^2\rho_\mathrm{2D}} 
- \frac{R_g^2}{R^2}\frac{\epsilon_0}{d^3\rho_\mathrm{2D}} (V_\mathrm{g}^\mathrm{DC})^2}
\label{eq:fresVg}
\end{equation}
for the resonance frequency as a function of $V_\mathrm{g}^\mathrm{DC}$. At $V_\mathrm{g}^\mathrm{DC} = 0~$V we get in agreement with Ref.~\cite{timoshenko1974SI}
\[
\omega_\mathrm{m}(0) = \frac{2.404}{R}\sqrt{\frac{Eh\epsilon}{\rho_\mathrm{2D}}}.
\]

\subsection{Harmonic oscillator model with effective mass}
It is instructive and useful to analyze the dynamics of the resonator by a harmonic oscillator model with an effective mass. 
From the total kinetic energy
\begin{equation}
E_\mathrm{kin} = \frac{1}{2}\rho_\mathrm{2D} 2\pi\int{r \xi_\mathrm{f}^2(r)dr} = \frac{1}{2} m_\mathrm{eff} \hat{z}^2
\end{equation}
we obtain for the effective mass
\begin{equation}
m_\mathrm{eff} = 0.27\rho_\mathrm{2D}\pi R^2,
\end{equation}
with
\[
2\pi\int_0^{R}{dr J_0\left(\frac{2.4}{R} r\right)r} = 2\pi\frac{R^2}{2.4^2}\int_0^{2.4}{dr'J_0(r')r'} = 0.27\pi R^2.
\]
We multiply all the terms of Eq.~\eqref{eq:localmotion} by $ J_0\left(\frac{2.4}{R} r\right)$ and integrate over the area.
As a result, we get the normalized equation of motion with higher order corrections for the capacitive force
\begin{eqnarray}
m_\mathrm{eff} \omega^2 \hat{z} &=& \left(0.271\pi R^2 2.4^2T -  0.271\frac{\epsilon_0 \pi R_\mathrm{g}^2 V_\mathrm{g}^2}{d^3}\right)\hat{z} \\
& & + 0.196 \frac{3 \epsilon_0 \pi R_\mathrm{g}^2 V_\mathrm{g}^2}{2d^4}\hat{z}^2 \notag\\
& & + 0.125 \frac{2 \epsilon_0 \pi R_\mathrm{g}^2 V_\mathrm{g}^2}{d^5}\hat{z}^3 +\ldots .\notag
\end{eqnarray}
Note that we obtain the same expression as Eq.~\ref{eq:fresVg} for the resonance frequency
\[
\omega_\mathrm{m}(V_\mathrm{g}^\mathrm{DC}) = \sqrt{\frac{4.92 Eh\epsilon}{m_\mathrm{eff}}
- \frac{0.271}{m_\mathrm{eff}}\frac{\epsilon_0\pi R_\mathrm{g}^2}{d^3} (V_\mathrm{g}^\mathrm{DC})^2}=\sqrt{\frac{2.4^2T}{R^2\rho_\mathrm{2D}} 
- \frac{R_\mathrm{g}^2}{R^2}\frac{\epsilon_0}{d^3\rho_\mathrm{2D}} (V_\mathrm{g}^\mathrm{DC})^2}.
\]

\subsection{Induced motion over capacitive drive}
First we are interested in the mechanical response under a weak electrostatic drive, such that nonlinear motional effects can be neglected:
\begin{equation}
m_\mathrm{eff}\ddot{z}(t)+\gamma_m m_\mathrm{eff}\dot{z}(t)+m_\mathrm{eff}\omega_m^2 z(t) = \hat{F}_\mathrm{d}\cos{(\omega_d t)}.
\end{equation}
The damping is characterized by the linewidth $\gamma_\mathrm{m} = \omega_\mathrm{m}/Q_\mathrm{m}$ with $Q_\mathrm{m}$ the quality factor of the mechanical resonator and $\omega_m$ the mechanical resonance frequency. The electrostatic drive amplitude is given by $\hat{F}_\mathrm{d}= \partial_z C_\mathrm{m}V_\mathrm{g}^\mathrm{DC}\sqrt{2}V_\mathrm{g}^\mathrm{AC}$, with $V_\mathrm{g}^\mathrm{AC}$ the root-mean-square amplitude of the drive voltage. Including the capacitive correction for the modeshape from above we have $\hat{F}_\mathrm{d} = 0.433\cdot\epsilon_0\pi R_\mathrm{g}^2/d^2 V_\mathrm{g}^\mathrm{DC}\sqrt{2}V_\mathrm{g}^\mathrm{AC}$.
With the ansatz $z(t)=\hat{z} e^{i\omega_\mathrm{d} t}$ we get
\begin{equation}
-m_\mathrm{eff}\omega_\mathrm{d}^2 \hat{z} + i\gamma_m m_\mathrm{eff}\omega_\mathrm{d} \hat{z} + m_\mathrm{eff}\omega_m^2 \hat{z} = \hat{F}_\mathrm{d}
\end{equation}
and hence for the amplitude
\begin{equation}
\hat{z}(\omega_\mathrm{d}) = \frac{F_\mathrm{d}/m_\mathrm{eff}}{\sqrt{(\omega_\mathrm{m}^2-\omega_\mathrm{d}^2)^2+\gamma_\mathrm{m}^2\omega_\mathrm{d}^2}}.
\label{eq:driven_lin}
\end{equation}
When driving at resonance $\omega_\mathrm{d}=\omega_\mathrm{m}$, we have
\begin{equation}
\hat{z}(\omega_\mathrm{m}) = \frac{F_\mathrm{d}}{m_\mathrm{eff}\gamma_m\omega_\mathrm{d}}.
\end{equation}

\subsection{Nonlinear lineshape}
We include the cubic nonlinear force in the driven equation of motion
\begin{equation}
m_\mathrm{eff}\ddot{z}(t)+i\gamma_m m_\mathrm{eff}\dot{z}(t)+m_\mathrm{eff}\omega_m^2 z(t) +\alpha_\mathrm{eff}z^3(t) = \hat{F}_\mathrm{d}\cos{(\omega_\mathrm{d} t)}.
\end{equation}
with $\alpha_\mathrm{eff}$ a constant.

In analogy to the linear lineshape in Eq.(\ref{eq:driven_lin}) we get for the amplitude of the motion
\begin{eqnarray*}
\hat{z}(\omega_\mathrm{d}) 
 &\approx& \frac{\hat{F}_\mathrm{d}/2m_\mathrm{eff}\omega_m^2}{\sqrt{\left(\frac{\omega_\mathrm{d}-\omega_m}{\omega_m}-\frac{3}{8}\frac{\alpha_\mathrm{eff}}{m_\mathrm{eff}\omega_m^2}\hat{z}_0^2\right)^2+(2Q)^{-2}}}\\
\end{eqnarray*}
in the limit of small oscillations where $\alpha_\mathrm{eff} \hat{z}^3 < k \hat{z}/Q_\mathrm{m}$ (see Ref.~\cite{lif08SI} Eq.~1.31a).
The onset of bistability is given by~\cite{lif08SI}
\begin{equation}
\hat{z}_\mathrm{crit} = \sqrt{\frac{8}{3\sqrt{3}}\frac{m_\mathrm{eff}\omega_m^2}{Q\alpha_\mathrm{eff}}} = 1.24\sqrt{\frac{m_\mathrm{eff}\omega_m^2}{Q\alpha_\mathrm{eff}}}.
\end{equation}
Thus the Duffing nonlinearity can be calculated from the critical deflection amplitude
\begin{equation}
\alpha_\mathrm{eff} = \frac{8}{3\sqrt{3}}\frac{m_\mathrm{eff}\omega_m^2}{Q_\mathrm{m}\hat{z}_\mathrm{crit}^2}.
\end{equation}

\subsection{Harmonic oscillator with nonlinear contributions and static displacement}
We consider quadratic and cubic nonlinear terms in the equation of motion (without dissipation and drive)
\begin{equation}
m_\mathrm{eff}\ddot{z}(t) = -m_\mathrm{eff}\omega_m^2 z(t) -\beta_0 z^2(t) -\alpha_0 z^3(t) + F_\mathrm{el},
\end{equation}
with $\beta_0$ and $\alpha_0$ two constants. 
With the separation ansatz $z(t)=z_\mathrm{s}+z_\mathrm{f}(t)$ we get
\begin{eqnarray*}
m_\mathrm{eff}\ddot{z}_\mathrm{f}(t) &=& -\left[m_\mathrm{eff}\omega_\mathrm{m,0}^2 z_\mathrm{s} +\beta_0 z_\mathrm{s}^2 + \alpha_0 z_\mathrm{s}^3 - \frac{1}{2}\partial_z C_\mathrm{m}(z_\mathrm{s})V_\mathrm{g}^2  \right]\\
& & -\underbrace{\left[m_\mathrm{eff}\omega_\mathrm{m,0}^2 +2 \beta_0 z_\mathrm{s} + 3\alpha_0 z_\mathrm{s}^2 - \frac{1}{2}\partial_z^2 C_\mathrm{m}(z_\mathrm{s})V_\mathrm{g}^2  \right]}_{k_\mathrm{tot}}z_\mathrm{f}(t)\\
& & -\underbrace{\left[\beta_0 + 3\alpha_0 z_\mathrm{s} - \frac{1}{4}\partial_z^3 C_\mathrm{m}(z_\mathrm{s})V_\mathrm{g}^2  \right]}_{\beta_\mathrm{tot}}z_\mathrm{f}^2(t)\\
& & -\underbrace{\left[\alpha_0 - \frac{1}{12}\partial_z^4 C_\mathrm{m}(z_\mathrm{s})V_\mathrm{g}^2  \right]}_{\alpha_\mathrm{tot}}z_\mathrm{f}^3(t)\\
\end{eqnarray*}

From the first bracket we can estimate the static displacement by neglecting the nonlinear contributions and assuming a similar deflection profile as the fundamental oscillation
\begin{eqnarray*}
z_\mathrm{s} &\approx& \frac{1}{2 m_\mathrm{eff}\omega_\mathrm{m,0}^2}\partial_z C_\mathrm{m} (V_\mathrm{g}^\mathrm{DC})^2\\
&\approx& \frac{0.433}{2 m_\mathrm{eff}\omega_\mathrm{m,0}^2} \frac{\epsilon_0 \pi R_g^2}{d^2}(V_\mathrm{g}^\mathrm{DC})^2\\
& = & \frac{\epsilon_0 R_\mathrm{g}^2}{7.21 T d^2}(V_\mathrm{g}^\mathrm{DC})^2.
\end{eqnarray*}
Compared to the result of the direct calculation with the static modeshape in Eq.~(\ref{eq:staticdispSI}) there is a small difference with a factor 7.21 instead of 8 in the denominator. 
For device $B$, it is possible to analyze the nonlinear contribution in $k_\mathrm{tot}$. For $z_\mathrm{s} = 17~$nm the nonlinear contribution equals the linear contribution as $3\alpha_0 z_\mathrm{s}^2 = m_\mathrm{eff}\omega_\mathrm{m}^2 = 1.6$~kg$\cdot$s$^{-2}$ with $\alpha_0 = 1.9\times 10^{15}$ kg$\cdot$s$^{-2}$m$^{-2}$. 

For small nonlinear amplitudes we transform the quadratic and cubic nonlinear terms in a single cubic term~\cite{nayfeh2008SI}
\begin{eqnarray*}
\alpha_\mathrm{eff} &\approx& \alpha_\mathrm{tot}-\frac{10}{9}\frac{\beta_\mathrm{tot}^2}{m_\mathrm{eff}\omega_m^2}\\
&\approx& \alpha_0 - \frac{1}{12}\partial_z^4 C_\mathrm{m}(V_\mathrm{g}^\mathrm{DC})^2-\frac{10}{9m_\mathrm{eff}\omega_m^2}\left(3\alpha z_\mathrm{s} - \frac{1}{4}\partial_z^3 C_\mathrm{m}(V_\mathrm{g}^\mathrm{DC})^2\right)^2\\
&\approx& \alpha_0 - \frac{1}{12}\partial_z^4 C_\mathrm{m}(V_\mathrm{g}^\mathrm{DC})^2 -\frac{10}{144 m_\mathrm{eff}\omega_m^2}\partial_z^3 C_\mathrm{m}^2(V_\mathrm{g}^\mathrm{DC})^4 -\frac{10}{m_\mathrm{eff}\omega_m^2}\alpha_0^2 z_\mathrm{s}^2.
\end{eqnarray*}
We assumed that $\partial_z^n C_\mathrm{m}(z_\mathrm{s}) \approx \partial_z^n C_\mathrm{m}(z_\mathrm{s}=0)$  and that $\beta$ is small (no symmetry breaking visible at small $V_\mathrm{g}^\mathrm{DC}$). With $\alpha_0 = 1.9\times 10^{15}$ kg s$^{-2}$m$^{-2}$ and $V_\mathrm{g} = 5~$V, the second and the third terms of the last equation are $\approx -6\times10^{12}$kg s$^{-2}$m$^{-2}$ and $\approx -6\times10^{11}$kg s$^{-2}$m$^{-2}$ respectively. This is much smaller than the fourth term ($\approx -4\times10^{15}$kg s$^{-2}$m$^{-2}$). Hence we can write
\begin{equation}
\alpha_\mathrm{eff} \approx \alpha_0 -\frac{10}{m_\mathrm{eff}\omega_m^2}\alpha_0^2 z_\mathrm{s}^2.
\end{equation}
The measured values for the Duffing nonlinearities are within the range of $\alpha_\mathrm{eff} = 1.74\cdot 10^{12}$~kg$\cdot$m$^{-2}$s$^{-2}$ to $7.16\cdot 10^{17}~$kg$\cdot$m$^{-2}$s$^{-2}$ observed in other graphene resonators~\cite{eichler11SI,song2011SI} and are compatible with the observation of intermediate strain.

\subsection{Heating of the mechanical resonator by large pump fields}
\begin{figure}[htb]
\centering
\includegraphics[width=0.8\linewidth]{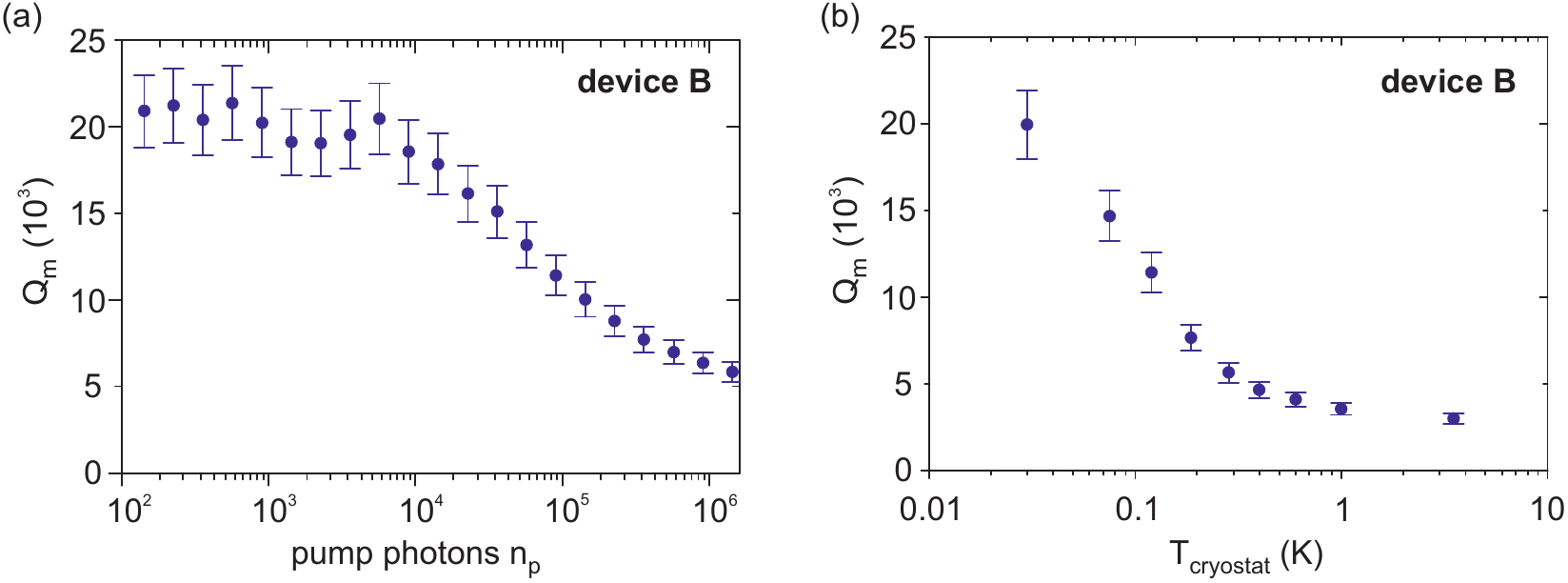}
\caption[Figure S2] {Heating of the mechanical mode by increasing the pump field in device $B$.
}
\label{f:figure S2}
\end{figure}
In Fig.~S3a we show a measurement of the quality factor in device $B$ as a function of the number of pump photons in the cavity. While the quality factor is roughly constant for $n_\mathrm{p} < 6000$, $Q_\mathrm{m}$ decreases for higher pump fields. Upon increasing the temperature of our cryostat $Q_\mathrm{m}$ also decreases, as shown in Fig.~S3b. From the comparison between the two figures we conclude that a pump power of $n_\mathrm{p} = 10^6$ has the same influence on the mechanical resonator as heating the cryostat to 200~mK. In order to minimize the heating it is beneficial to reduce the resistance of the graphene and to improve the heat flow away from the mechanical resonator.


%


\end{document}